\documentstyle[prl,aps,epsfig]{revtex}
\begin{document}

\def\lsim{\mathrel{\rlap{
\lower4pt\hbox{\hskip-3pt$\sim$}}
    \raise1pt\hbox{$<$}}}     
\def\gsim{\mathrel{\rlap{
\lower4pt\hbox{\hskip-3pt$\sim$}}
    \raise1pt\hbox{$>$}}}     

\draft

 \twocolumn[\hsize\textwidth\columnwidth\hsize  
 \csname @twocolumnfalse\endcsname              

\title{Shell Model Embedded in the Continuum for Binding Systematics in
Neutron-Rich Isotopes of Oxygen and Fluor}

\author{Y. Luo\dag, J. Oko{\l}owicz\dag\S, M. P{\l}oszajczak\dag ~and 
N. Michel\dag}
\address{\dag\ Grand Acc\'{e}l\'{e}rateur National d'Ions Lourds (GANIL),
CEA/DSM -- CNRS/IN2P3, BP 5027, F-14076 Caen Cedex 05, France}
\address{\S\ Institute of Nuclear Physics, Radzikowskiego 152,
PL - 31342 Krakow, Poland}

\maketitle

\begin{abstract}
Continuum coupling correction to binding energies 
in the neutron rich oxygen and fluorine isotopes is studied using 
the Shell Model Embedded in the Continuum. We discuss the
importance of different effects, such as the position of one-neutron 
emission threshold, the effective interaction or the number of valence 
particles on the magnitude of this correction.
\end{abstract}


 ]  

\narrowtext

Nuclear structure of light neutron-rich nuclei has attracted much attention
because of unusual spatial features of these nuclei \cite{tanihata1}
and the isospin effects in the magic numbers \cite{tanihata}. 
In this respect, the systematics of neutron
separation energies $S_n$ and interaction cross-section in $sd$ shell region
seem to indicate the disappearance of $N=20$ closure and the appearance of new shell
closure at $N=16$ in neutron-rich nuclei with $T_Z\geq 3$ close to the 
drip-line \cite{tanihata}. Another astonishing property is a large change of 
the neutron drip-line by adding one proton to the oxygen core, as indicate
the instability of oxygen isotopes for $N>16$ \cite{exp1,exp2} 
and the stability of $^{31}$F with $N=22$ \cite{exp2}.
Theoretical description of the low-lying states properties in these nuclei
requires taking into account the coupling between discrete and continuum
correlated states \cite{naza1,bnop1}. This aspect is particularly important 
near drip-line where 
one has to use both structure and reaction data to understand 
basic shell structure and dynamics of these nuclei \cite{tanihata,bnop2}. 
In weakly bound light nuclei, not only coupling of bound states to continuum 
is strong but also many-body correlations should be treated rigorously.
Both the continuum  coupling
and the realistic configuration mixing can be taken into account 
in the Shell Model Embedded in the Continuum (SMEC) \cite{bnop1}. 
This approach provides a 
unified description of diverse nuclear characteristics such as the energy 
spectra, including nucleon emission widths and electromagnetic 
transition probabilities,
and the reactions involving one-nucleon in the continuum
\cite{bnop1,bnop2}. 
In this work, we analyze in the Shell Model (SM) framework the continuum
correction to binding energies and neutron separation energies. 

The detailed description of SMEC formalism has been given elsewhere 
\cite{bnop1,bnop2}. Separation of (quasi-) bound ($Q$ subspace) 
and scattering ($P$ subspace) states is achieved using the projection 
operator technique \cite{bartz3}. $P$ contains asymptotic channels 
made of $(A-1)$-particle localized states and one nucleon in the scattering 
state. Localized many-body states which are
build up by the bound and resonance single-particle (s.p.) wave functions, are
included in $Q$. They are obtained by solving standard multiconfigurational 
SM problem for the Hamiltonian $H_{QQ}$. 
The residual coupling $H_{PQ}$ between states 
in $Q$ and $P$ is given by \cite{bnop1} :
$V_{12} = -V_{12}^{(0)}[\alpha + \beta P_{12}^{\sigma}]
\delta({\bf r}_1 - {\bf r}_2)$, where $\alpha+\beta=1$ and $P_{12}^{\sigma}$ 
is the spin exchange operator. The effective SM Hamiltonian including the
coupling to the continuum is energy-dependent : 
$H_{QQ}^{\rm eff}(E)=H_{QQ}+H_{QP}G_P^{(+)}(E)H_{PQ}$, where 
$G_P^{(+)}(E)$ is a Green function for the motion of single nucleon 
in the $P$ subspace. $H_{QQ}^{\rm eff}$ is a complex-symmetric
matrix above $E^{\rm (thr)}$ and Hermitian below it.   
The energy scale is settled by the one-nucleon emission
threshold $E^{\rm (thr)}$ \cite{bnop1}. Radial s.p.\ wave functions in $Q$ 
and the scattering wave functions in $P$ are generated in a self-consistent
procedure, starting with the average potential of Woods-Saxon type
with the spin-orbit and Coulomb parts included, and taking into account the
residual coupling. This procedure yields new orthonormalized wave functions in 
$Q$ and $P$ and new self-consistent potentials for each many-body state in $Q$.

In the present studies we use the full $sd$ valence 
space for $N<20$ and the full $pf$ shell for $N>20$. 
For the effective interaction in $H_{QQ}$ we take USD
Hamiltonian for the $sd$ shell \cite{wildenthal} and the KB$^{'}$
interaction for the $pf$ shell \cite{strasbourg}. The cross-shell interaction
is the $G$-matrix \cite{lks}. In the calculation of binding
energies from the nuclear energies in SMEC and SM, we proceed as in Refs. \cite{reta}.

The ground state (g.s.) continuum coupling correction in SMEC is calculated as :
$E_{\rm corr}=<\Phi_{g.s.}|H_{QQ}^{\rm eff}-H_{QQ}|\Phi_{g.s.}>$, 
the g.s. wave function in parent nucleus $(N,Z)$ is coupled to
different channel wave functions, which are determined by the motion of 
an unbound neutron  relative to the daughter nucleus 
$(N-1,Z)$ in a certain SM state $\Phi_i^{(N-1)}$. 
The magnitude of coupling matrix elements 
vary depending on the structure of SM wave function $\Phi_i^{(N-1)}$ and the
value of one-neutron emission threshold $E_n^{\rm (thr)}$.
The individual contributions for different channel couplings 
tend to decrease with
increasing excitation energy of a SM state $\Phi_i^{(N-1)}$ due to a growing mismatch of
both quantum numbers 
and radial properties of wave functions in parent and 
daughter nuclei \cite{rotter}. This decrease is exponential on average. 
In the present studies, we include {\it all} asymptotic channels 
composed of SM states; 
their number vary from 3 in $^{28}$O to 1837 in $^{31}$F. 
Even though in most nuclei these channels are closed, nevertheless
the coupling to them modifies the binding energy
depending both on $E_n^{\rm (thr)}$ and the structure of the g.s. wave function.
\begin{figure}
\vspace{-0.5cm}
\psfig{file=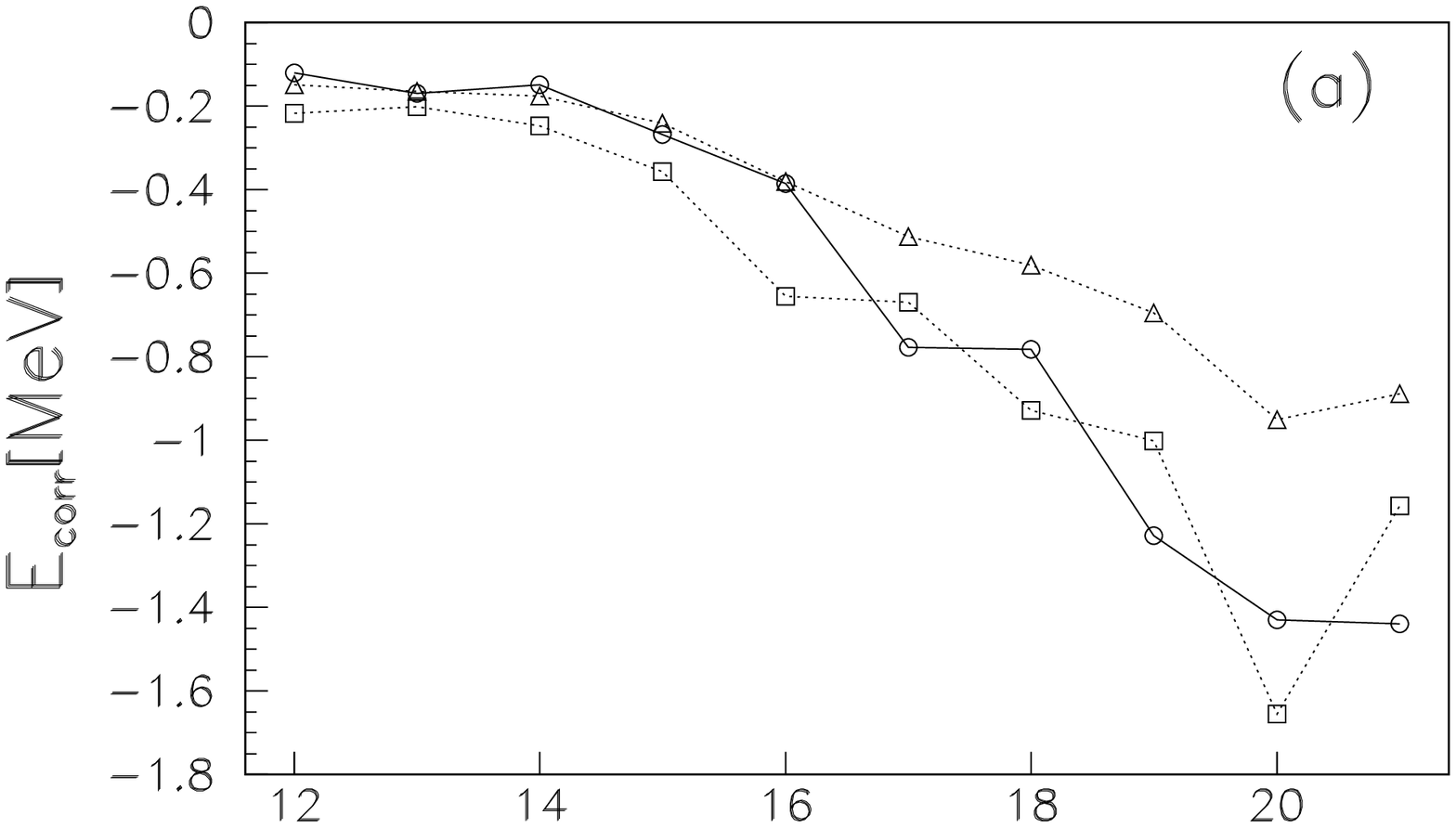,height=7.5cm,angle=00}
\vspace{-3.5cm}
\end{figure}
\begin{figure}
\vspace{-0.5cm}
\psfig{file=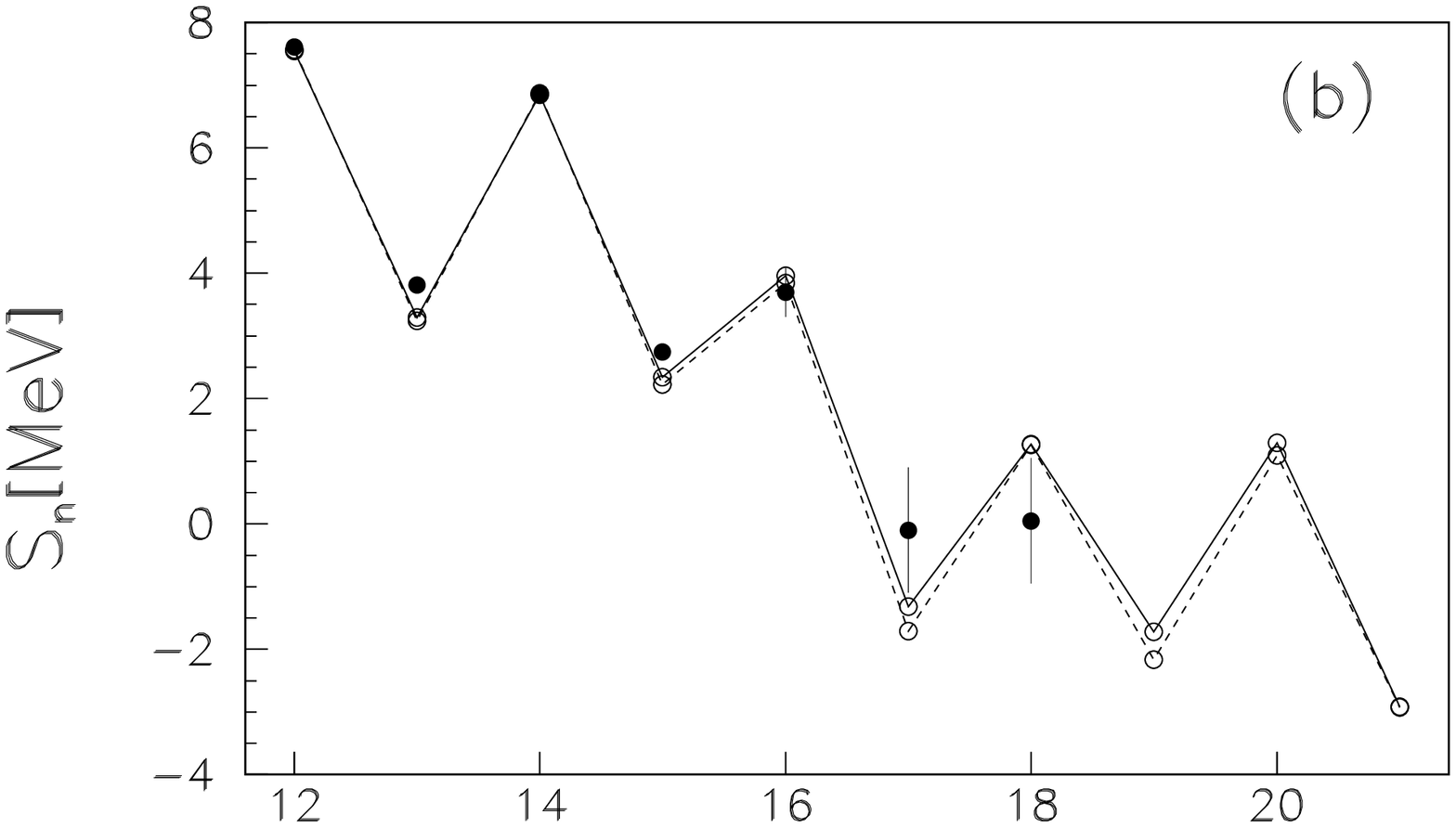,height=7.5cm,angle=00}
\vspace{-3.5cm}
\end{figure}
\begin{figure}
\vspace{-0.5cm}
\psfig{file=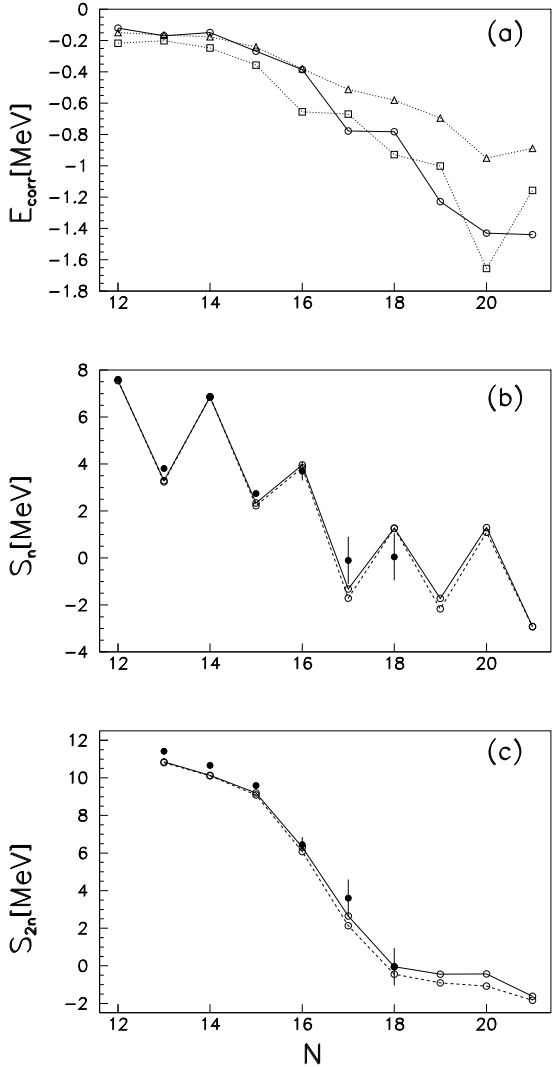,height=7.5cm,angle=00}
\vspace{-2cm}
\caption{The systematics of $0\hbar\omega$ SMEC correction in the 
oxygen isotopes. (a) Neutron number dependence of the SMEC energy
correction to the SM g.s. energy. The solid line is obtained for one-neutron 
emission threshold $E_n^{\rm (thr)}$ calculated in SMEC for each nucleus. 
The dotted line with squares and triangles is obtained for 
$E_n^{\rm (thr)}$ which is fixed arbitrarily at 0 and 4 MeV, respectively. 
(b) One-neutron separation energies $S_n$ calculated
in SM (dashed line) and in SMEC (solid line) are compared with the data (closed
circles with the error bars)
\protect\cite{audi}. 
(c) Two-neutron separation energies $S_{2n}$ in SM and SMEC. Symbols are the
same as in (b). 
}
\end{figure}

The choice of the initial average 
potential follows a procedure described in Ref. \cite{bnop2}. 
We demand, that the self-consistent potentials yield 
the energies of s.p. states at the same position as in the average potential
with Woods-Saxon central part of radius $R_0=1.27A^{1/3}$ fm, 
depth $V_0(N,Z)=-51.1 [1\mp0.69(N-Z)/(N+Z)]$ MeV ($-/+$ stands for neutron/proton
potential), diffuseness $a=0.67$ fm. The spin-orbit potential is 
$ V_{\rm so} {\lambdabar}_{\pi}^2 (2{\bf l}\cdot{\bf s})r^{-1}d{f}(r)/dr$
with ${\lambdabar}_{\pi}^2 = 2\,$fm$^2$, $V_{\rm so}(N,Z)=0.131V_0(N,Z)$
MeV and radial formfactor $f(r)$ of the central potential. The Coulomb potential is calculated for
the uniformly charged sphere with radius $R_0$.  

Fig. 1 shows results for oxygen isotopes. In our $0\hbar\omega$ description,
the continuum coupling in these nuclei contains only
neutron-neutron ($T=1$) part. For the strength 
$V_{0}^{(nn)}\equiv V_{12}^{(0)}(\alpha-\beta)=414$ MeV$\cdot$fm$^3$, 
one obtains a good overall agreement with the spectra of oxygen isotopes. 
For example, 
SMEC calculation in $^{24}$O yields the first excited state $2_1^+$ 
above the experimental $E_n^{\rm (thr)}$, in agreement with the 
data \cite{stanoiu}. The $nn$-continuum coupling modifies $T=1$ monopole 
terms \cite{zuker} of the SM interaction. A good agreement for the binding systematics 
in oxygen chain can be found provided the $T=1$ part of two-body interaction 
is modified as follows : $\delta V_{0d_{5/2},1s_{1/2}}^{T=1}=-0.05$ MeV, 
$\delta V_{1s_{1/2},0d_{3/2}}^{T=1}=+0.25$ MeV,  
$\delta V_{0d_{3/2},0d_{3/2}}^{T=1}=-0.35$ MeV. 

Fig. 1a shows the neutron number dependence of 
$E_{\rm corr}$ for (i) $E_n^{\rm (thr)}$ of SMEC 
(solid line), and for (ii) $E_n^{\rm (thr)}$ fixed arbitrarily at 0 and 4 MeV
(the dotted lines with open squares and triangles, respectively).  
The $N$-dependence of $E_{\rm corr}$ in these three cases
exhibits approximately quadratic dependence on the number of valence neutrons, 
which is usual for a monopole Hamiltonian \cite{zuker}. 
Deviations from this dependence are characteristic 
of the continuum coupling. For 
$E_n^{\rm (thr)}=0$ (a limit of one-neutron drip-line), 
one can see an odd-even staggering (OES) of $E_{\rm corr}$ around an average 
$N$-dependence. A blocking of the virtual scattering to the particle continuum 
by an odd nucleon diminishes the continuum correction to the
binding energy  of broken-pair g.s. of odd-$N$ nuclei. 
This "drip-line effect", which is seen also in the coordinate-space
Hartree-Fock-Bogolyubov approach \cite{hfb}, is restricted to a narrow window of
excitation energies around $E_n^{\rm (thr)}=0$. It vanishes if $E_{\rm corr}$
is calculated for $E_n^{\rm (thr)}=4$ MeV (see the dotted line with triangles in
Fig. 1a). 

If $E_n^{\rm (thr)}$ is calculated in SMEC  
(the solid line), one finds an opposite OES effect with enhanced $E_{\rm corr}$ for 
odd-$N$ isotopes. This is due to $E_n^{\rm (thr)}$ which is lower
in odd-$N$ isotopes than in neighbouring even-$N$ isotopes. 
This threshold induced OES effect
leads to the attenuation of OES of binding energies. An enhanced modification 
of one-neutron separation energies $S_n$ is seen 
for odd-$N$ oxygen isotopes (see Fig. 1b). In general, 
the effect of continuum coupling on one- and 
two-neutron separation energies is seen mainly for $S_n \lsim 3$ MeV, i.e.
close to the one-neutron drip-line.

In fluorine isotopes, both neutron-neutron ($T=1$) and neutron-proton ($T=0,1$) 
couplings between states in $Q$ and $P$ are present. 
The $nn$-coupling has been adjusted in the oxygen isotope 
chain and its value is kept unchanged. The
optimal value of the strength of the $np$-continuum coupling : $V_0^{(np)}\equiv
V_{12}^{(0)}(\alpha+\beta)$, varies 
from a standard value $V_0^{(np)}\simeq 2V_0^{(nn)}$ for well-bound isotopes at
the valley of stability to 
$V_0^{(np)}\simeq (1/2)V_0^{(nn)}$ near the neutron drip-line.  
This coupling modifies strongly 
$T=0$ monopole terms of the SM Hamiltonian. For 
$V_0^{(np)}\simeq (1/2)V_0^{(nn)}$, good agreement for
$S_{n}$, $S_{2n}$ and OES of binding energies \cite{satula} :
\begin{eqnarray}
\label{eq1}
\Delta^{(3)}(N)=((-1)^N/2)[E(N+1)-2E(N)+E(N-1)] 
\end{eqnarray}
is found provided the $T=0$ two-body interaction is modified as follows :  
$\delta V_{0d_{5/2},0d_{3/2}}^{T=0}=+0.6$ MeV, 
$\delta V_{0d_{5/2},0f_{7/2}}^{T=0}=+0.9$ MeV.  

Fig. 2a shows the neutron
number dependence of $E_{\rm corr}$ for fluorine isotopes, which is calculated 
for $V_0^{(np)} = (1/2)V_0^{(nn)}$. $E_{\rm corr}$ depends both on $|N_p-N_n|$ and
$(N_p+N_n)$ (compare Figs. 1a and 2a), where $N_p$, $N_n$ are the number of 
protons and neutrons in the valence space. Three cases shown in Fig. 2a 
can be directly compared with those shown in Fig. 1a. 
For $E_n^{\rm (thr)}=4$ MeV, one can see the OES of $E_{\rm corr}$ 
which is absent in the oxygen chain. 
This is the salient feature of the
$np$-coupling. $E_{\rm corr}$ in odd-odd isotopes is increased
as compared to the neighboring odd-even ones. The size of $E_{\rm corr}$ in this
case is weakly dependent on the precise 
value of $E_n^{\rm (thr)}$. Qualitatively similar effect can be seen also at the 
neutron drip-line (see the results for $E_n^{\rm (thr)}=0$ in Fig. 2a) but now the 
OES due to the $np$-coupling is attenuated by an opposite 
staggering due to the $nn$-coupling (see also 
Fig. 1a). If $E_n^{\rm (thr)}$ is calculated in SMEC for each nucleus 
(the solid line), the OES of $E_{\rm corr}$ around a smooth $N$-dependence is 
enhanced due to combined effects of the $np$-continuum coupling, 
which weakly depends on $E_n^{\rm (thr)}$, and the $nn$-continuum coupling, 
which closely follows the OES of 
$E_n^{\rm (thr)}$. These two effects act 'in phase', hence contributing
an additional binding for odd-$N$ nuclei. 

A strong sensitivity of $E_{\rm corr}$ in fluorine chain to the value of the strength 
$V_0^{(np)}$ can be seen in Figs. 2b and 2c which show the neutron number
dependence of $S_n$ and $S_{2n}$. The dashed line shows the SM results, whereas
the solid line gives SMEC results for $V_0^{(np)} = (1/2)V_0^{(nn)}$. The
agreement of SMEC results with the experimental data is excellent for
neutron-rich nuclei. The dotted line in Figs. 2b
and 2c corresponds to SMEC calculation with $V_0^{(np)} = 2V_0^{(nn)}$. 
For this standard value, one finds a somewhat better
agreement with the experimental data for
light fluorine isotopes. On the contrary, description of neutron-rich
nuclei is poor. 

A useful indicator of OES is $\Delta^{(3)}(N)$ (Eq. \ref{eq1}). 
Fig. 3 compares experimental data for $\Delta^{(3)}(N)$ in 
fluorine isotopes with those calculated in SMEC and in SM. 
SMEC results correspond to : $V_0^{(np)} = (1/2)V_0^{(nn)}$ (the solid line) 
and $V_0^{(np)} = 2V_0^{(nn)}$ (the dotted line). SM results have been
obtained using two different SM interactions : (i) with the modified $T=0,1$ 
monopole terms (the dashed line), as discussed above, and (ii) without these 
modifications (the dashed-dotted line). One can see that the SM calculations
overestimate systematically the OES effect, in particular
close to the neutron drip-line. The $np$-continuum coupling 
strongly attenuates the OES and may even wash it out 
for values of ratio $V_0^{(np)}/V_0^{(nn)}$, close to the accepted value 
$\gsim 2$ in nuclei from the valley of stability. 
\begin{figure}
\vspace{-0.5cm}
\psfig{file=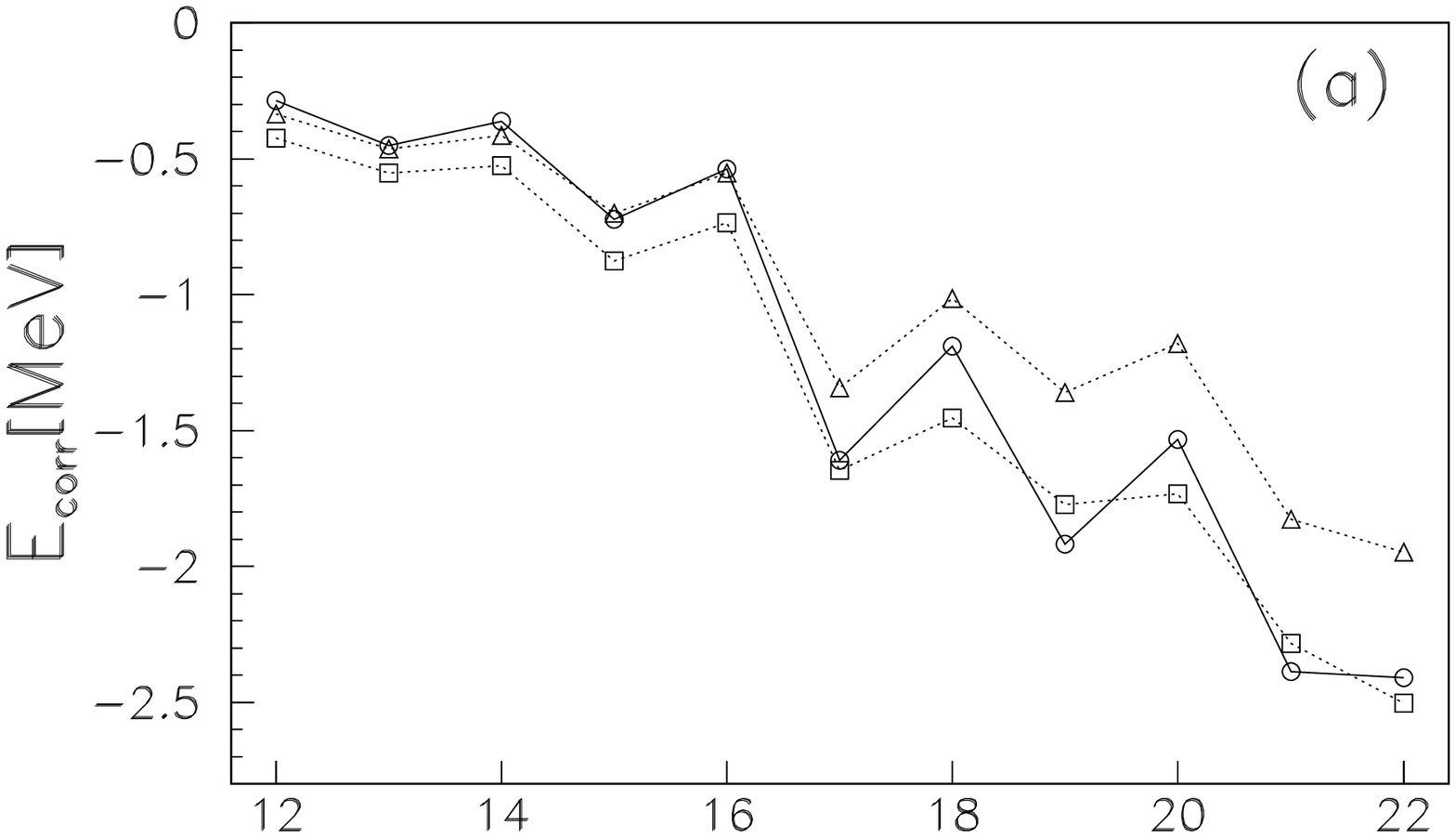,height=7.5cm,angle=00}
\vspace{-3.5cm}
\end{figure}
\begin{figure}
\vspace{-0.5cm}
\psfig{file=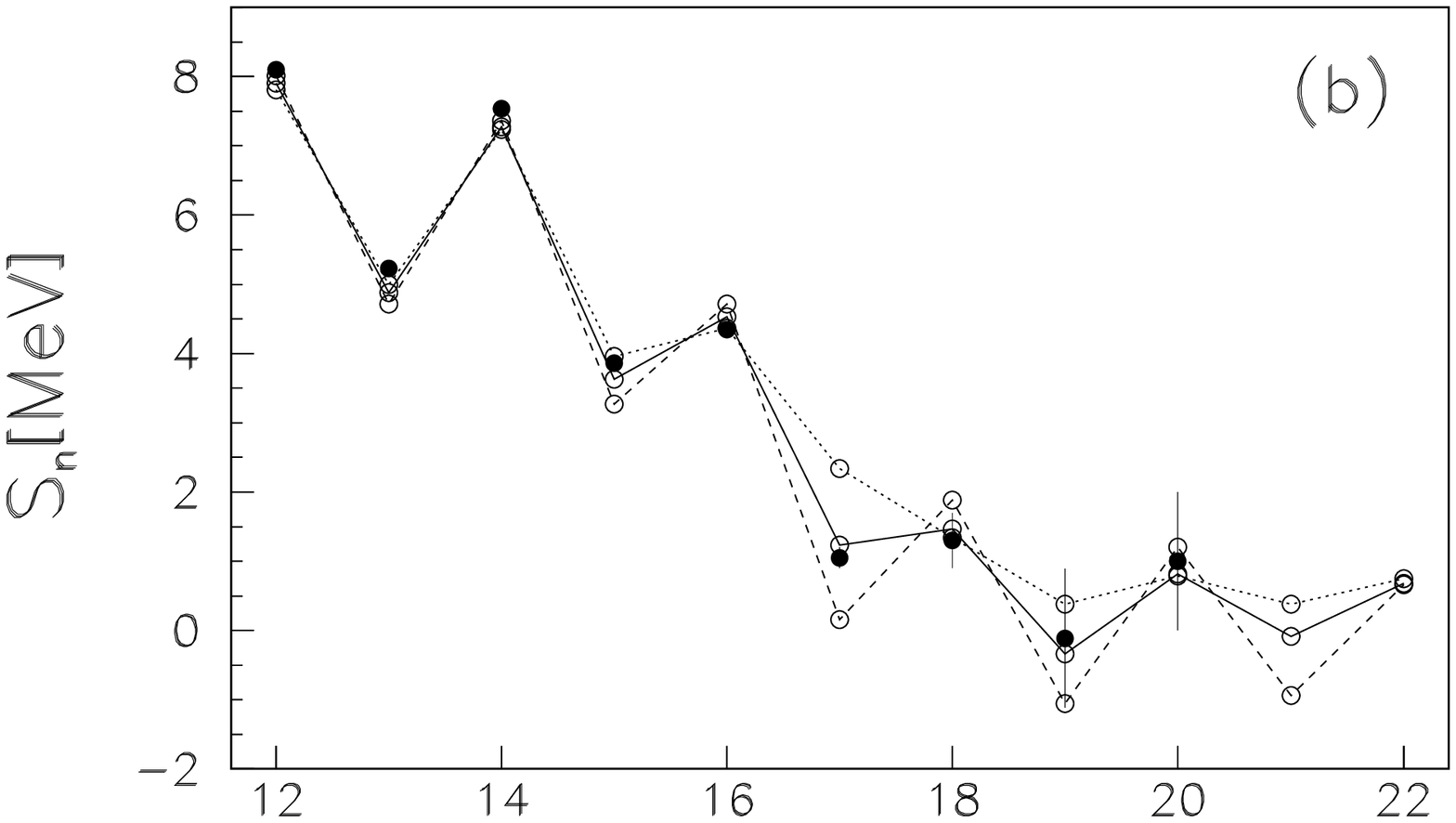,height=7.5cm,angle=00}
\vspace{-3.5cm}
\end{figure}
\begin{figure}
\vspace{-0.5cm}
\psfig{file=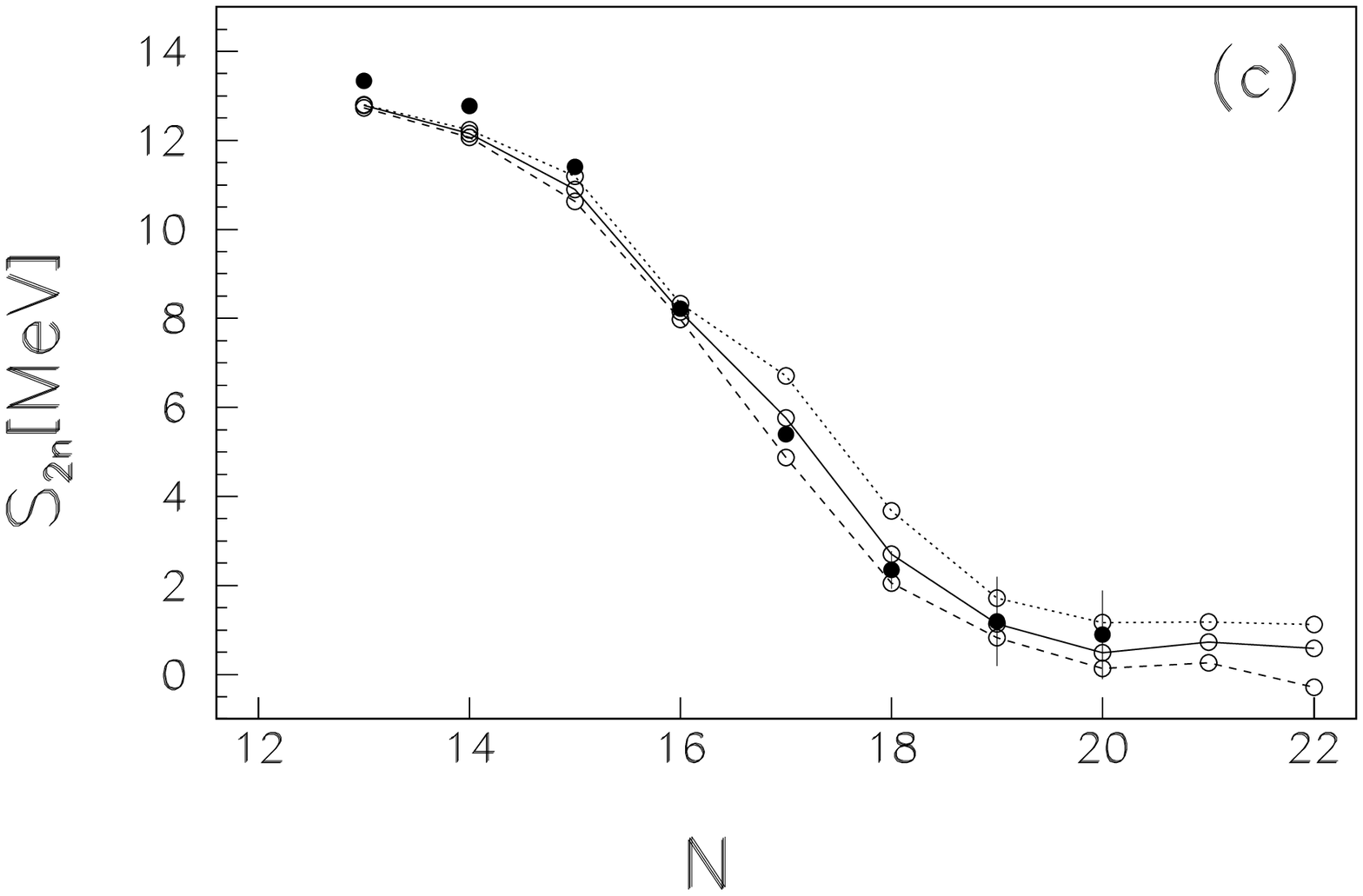,height=7.5cm,angle=00}
\vspace{-2cm}
\caption{The same as in Fig. 1 but for the fluorine isotopes. SMEC calculations
shown by the solid line are performed for the strengths 
of the continuum coupling : $V_{0}^{(nn)}=414$ MeV$\cdot$fm$^3$ and
$V_0^{(np)} = (1/2)V_0^{(nn)}$. The dotted lines in (b) and (c) have been
obtained for $V_0^{(np)} = 2V_0^{(nn)}$. The closed
circles with error bars shows the data \protect\cite{audi}.}
\end{figure}

In summary, we have performed first large-scale SM calculations including the
coupling to the continuum states which extend
from the valley of stability up to the neutron drip-line. Our analysis
demonstrates an important renormalization of the effective SM Hamiltonian 
by the continuum coupling; the principal effects of this renormalization 
depend on the position of the one-neutron emission threshold $E_n^{\rm (thr)}$, 
on isospin $T$, its projection $T_Z$ and on the number of valence particles. 
Certain features of this 
renormalization can be simulated in SM studies by a suitable shifts 
of the two-body matrix elements. The dependence of $E_{\rm corr}$ on both 
$E_n^{\rm (thr)}$ and the number of valence neutrons $N_n$ and protons $N_p$
implies, however, that the two-body monopoles in SM studies
must contain the additional terms depending on $E_n^{\rm (thr)}$, $|N_p-N_n|$, and
$(N_p+N_n)$, which become important near the drip line. 
\begin{figure}
\vspace{-0.5cm}
\psfig{file=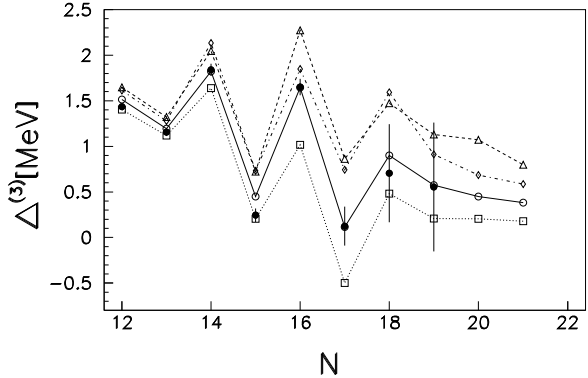,height=8cm,angle=00}
\vspace{-2cm}
\caption{The OES of binding energies (\protect\ref{eq1}) 
for SM (the dashed line) and SMEC with different ratios of $np$- and $nn$-
continuum coupling strengths :  $V_0^{(np)} = (1/2)V_0^{(nn)}$ 
(the solid line) and $V_0^{(np)} = 2V_0^{(nn)}$ (the dotted line). 
The dashed-dotted line with open diamonds, shows SM results  
for the USD+KB$^{'}$ interaction {\it without} the modification of $T=0,1$ 
monopole terms discussed in the text. The closed points with error bars shows
the data \protect\cite{audi}.}
\end{figure}

We have found an important attenuation of the OES of binding
energies close to the neutron drip-line by the 
$np$-continuum coupling. This attenuation does not mean that the
neutron pairing correlations become weaker 
(see Fig. 1a) but that the $np$-continuum coupling 
'falsifies' $\Delta^{(3)}(N)$ indicator of neutron pairing. 

The analysis of experimental data in SMEC indicates a significant 
attenuation of $np$-continuum coupling in nuclei with large neutron excess. 
A strong sensitivity of $S_n(N)$ and $\Delta^{(3)}(N)$ 
to the value of $V_0^{(np)}/V_0^{(nn)}$ allows to
estimate this attenuation quantitatively. 

The continuum coupling correction to energy of a SM state depends strongly on 
the position of this state with respect to $E_n^{\rm (thr)}$. 
This dependence induce downward 
shifts of excitation energy for SM states
close to $E_n^{\rm (thr)}$. The $T_Z-$dependence of these shifts (see Figs. 1a 
and 2a) implies that these SM states (e.g. intruder states)
, even for a constant value of one-neutron emission threshold,
will be pushed down by an amount which depends on $T_Z$.

The coupling to the two-neutron continuum has not been included in the present
analysis. This coupling is expected to be less important if $S_n < S_{2n}$,
i.e. for $N < 18$ in O- and F-isotope chains. Since $E_{2n}^{\rm (thr)}$ is a smooth
function of $N$ and, moreover, varies slowly near the neutron drip-line,
therefore we expect that the coupling to the 
two-neutron continuum produces a smaller effect on the OES of binding energies
than the coupling to the one-neutron continuum. One should keep in mind,
however, that additional attractive correlations between nucleons 
in the scattering continuum are responsible for the appearance of 
"Borromean systems" close to the one-neutron drip line \cite{gamow}.

\vskip 0.3truecm

We thank W. Nazarewicz for valuable suggestions and  F. Nowacki for useful
discussions.


\end{document}